\documentclass[12pt,preprint]{aastex}

\usepackage{lscape}
\begin{document}

\title{Very Low Mass Stars and Brown Dwarfs in Taurus-Auriga}

\author{Russel J. White\altaffilmark{1,2} and Gibor Basri\altaffilmark{3} }

\altaffiltext{1}{McDonald Observatory, R.L.M. Hall 15.308, Austin, TX
78712-1083}

\altaffiltext{2}{Present Address: Department of Astronomy, California
Institute of Technology, MS 105-24, Pasadena, CA 91125}

\altaffiltext{3}{Department of Astronomy, University of California at
Berkeley, Berkeley, CA 94720-3411}

\begin{abstract}

We present high resolution optical spectra obtained with the HIRES
spectrograph on the W. M. Keck I telescope of seven low mass T Tauri stars
and brown dwarfs (LMTTs) in Taurus-Auriga.  The observed Li I 6708 \AA\,
absorption, low surface gravity signatures, and radial velocities confirm
that all are members of the Taurus star forming region; no new
spectroscopic binaries are identified.  Four of the seven targets observed
appear to be T Tauri brown dwarfs.  Of particular interest is the
previously classified "continuum T Tauri star" GM Tau, which has a spectral
type of M6.5 and a mass just below the stellar/substellar boundary.

These spectra, in combination with previous high resolution spectra of
LMTTs, are used to understand the formation and early evolution of objects
in Taurus-Auriga with masses near and below the stellar/substellar
boundary.  None of the LMTTs in Taurus are rapidly rotating (vsin$i$ $<$ 30
km/s), unlike low mass objects in Orion.  Many of the slowly rotating,
non-accreting stars and brown dwarfs exhibit prominent H$\alpha$ emission
(equivalent widths of 3 - 36 \AA), indicative of active chromospheres.  We
demonstrate empirically that the full-width at 10\% of the H$\alpha$
emission profile peak is a more practical and possibly more accurate
indicator of accretion than either
the equivalent width of H$\alpha$ or optical veiling: 10\%-widths $> 270$
km/s are classical T Tauri stars (i.e. accreting), independent of stellar
spectral type.  Although LMTTs can have accretion rates comparable to that
of more typical, higher-mass T Tauri stars (e.g. K7-M0 spectral types), the
average mass accretion rate appears to decrease with decreasing mass.  A
functional form of $\dot{M} \propto M$ is consistent with the available
data, but the dependence is difficult to establish because of both
selection biases in observed samples, and the decreasing frequency of
active accretion disks at low masses (M $<$ 0.2 M$_\odot$).  The diminished
frequency of accretion disks for LMTTs, in conjunction with their lower, on
average, mass accretion rates, implies that they are formed with less
massive disks than higher-mass T Tauri stars.  The radial velocities,
circumstellar properties and known binaries do not support the suggestion
that many of the lowest mass members of Taurus have been ejected from
higher stellar density regions within the cloud.  Instead, LMTTs appear to
have formed and are evolving in the same way as higher-mass T Tauri stars,
but with smaller disks and shorter disk lifetimes.

\end{abstract}


\newpage

\section{Introduction}

Theories of star and planet formation have suggested that stars (M
$\gtrsim$ 0.075 M$_\odot$) form from the dynamical collapse of a cloud
core, while planets
(M $\lesssim$ 0.013 M$_\odot$) form via the accretional coagulation of
material in a circumstellar disk \citep[e.g.][]{boss89}.  The
intermediate mass at which the dominant formation mechanism changes from
collapse to coagulation within a disk, however, is not known.  Brown dwarfs
are objects with masses intermediate between those of stars and planets.
The formation process of brown dwarfs, therefore, offers an important
mass-link between star and planet formation.  Unfortunately, little is
known about the formation of brown dwarfs, or even of the the lowest mass
stars (M $< 0.2$ M$_\odot$), because of the difficulty in determining their
basic properties (e.g. mass accretion rates, rotational velocities,
spectroscopic binarity) at young ages.  Accurately determining these
properties generally requires high resolution spectroscopy.  At the
distances of the nearest regions of star formation, however, the lowest
mass stars and brown dwarfs are too faint ($R_c > 14$ mag) to have been
included in previous high spectral resolution surveys \citep[e.g.][]{bb90,
heg95}.

In order to ascertain the basic properties of young low mass stars and brown
dwarfs, we have carried out a high resolution spectroscopic study of mid-M
spectral type (and presumably low mass) T Tauri stars in the nearby 
Taurus-Auriga star forming region \citep[D $\sim$ 140 pc;][]{bertout99}.
We use these spectra to confirm the T Tauri classification of the sample
and to estimate effective temperatures from which mass estimates are
derived.  We then investigate the mass dependence of rotation and mass
accretion across the stellar/substellar boundary.  The results
are used to understand the formation of the lowest mass stars and brown
dwarfs in Taurus-Auriga.

\section{T Tauri Sample, Observations, and Data Reduction}

The observational goal of this study was to obtain high resolution spectra
of low mass T Tauri stars and brown dwarfs (LMTTs) in the star forming
region Taurus-Auriga.  At the time the observations were conducted (Dec
1999), there were roughly 30 LMTTs known in Taurus with spectral
types M4 or later \citep[M $\lesssim 0.3$ M$_\odot$;][]{bcah98}, based on
low resolution spectra \citep{briceno93, briceno98, gomez92, luhman98,
hss94, hb88, magazzu91}, with the coolest confirmed member of Taurus at a
spectral type of M7.5\footnote{Several cooler candidate Taurus members have
been proposed \citep{itoh99, rh99}, but are yet unconfirmed.  Since these
observations were conducted, several new low mass members have been
identified \citep{martin01, briceno02}}.  Within Taurus, there are also
several T Tauri stars whose optical spectra at low resolution are too
heavily veiled to detect the photospheric features needed for spectral
classification \citep[CIDA-1, GM Tau, GN Tau, Haro 6-13, FT
Tau;][]{briceno93, hb88}.  These stars are typically referred to as
'continuum' stars, although their faint optical emission ($m_v$ typically
$\gtrsim$ 17 mag) suggests that they could be of mid- or late-M spectral
type.  These stars were included as possible targets so as not to bias
against LMTTs experiencing high accretion.

From this possible source list, we observed the 4 mid-M LMTTs MHO-4, MHO-5,
MHO-9, and V410 X-ray 3, and the 3 continuum stars GM Tau, GN Tau
and CIDA-1.  High resolution spectra of these 7 target stars were obtained
on 1999 Dec 4 with the Keck I telescope using the High Resolution Echelle
Spectrometer \citep{vogt94}.  With the 1\farcs15 slit (HIRES Decker 'D1'),
the instrument yielded 15 spectral orders from 6390 - 8700 \AA\, (with gaps
between the orders) with a 2-binned-pixel (4 CCD pixels) spectral
resolution of R $\approx$ 33,000.  Seven mid- and late-M giant stars and
the M6 dwarf Gl 406 were also observed
on the same night with this setup for spectral comparison.  The properties
of these standard stars along with others used in the analysis are
summarized in the Appendix.  The data were reduced in a standard
fashion for echelle data using IDL routines as described in \citet{basri00}.
In the subsequent analysis, we include the high spectral resolution
observation of GG Tau Ba (M6), GG Tau Bb (M7.5), and UX Tau C (M5)
obtained with HIRES and reported in \citet{wgrs99} and \citet{bm95}.

We note that GN Tau is a 0\farcs33 binary star with components of
comparable brightness at optical and infrared wavelengths \citep{simon96,
wg01}.  The other 6 LMTTs observed here are known to be
speckle-singles based on high spatial resolution speckle observations using
the Keck telescope \citep{gws02}.  GG Tau Ba, GG Tau Bb, and UX Tau C are
members of wide ($>1\farcs4 = 200$ AU) multiple systems \citep{wg01}.

\section{Results}

\subsection{Spectroscopic Properties}

All 10 targets exhibit Li I 6708 \AA\, absorption and prominent
H$\alpha$ emission, distinguishing characteristics of T Tauri stars.
Equivalent widths (EW) of these features are listed in Table 1.
In this section, the spectra are used to determine stellar spectral types,
radial velocities, projected rotation velocities ($v$sin$i$), continuum
excesses, H$\alpha$ emission profiles, and relative surface gravities.

\subsubsection{Spectral Types}

Our goal in spectral type classification is to determine what stars of
known spectral type best represent the photospheric emission of the
observed M-type LMTTs.  Historically, the spectra of T Tauri stars, with
spectral types of K7-M0, have been successfully matched by the spectra
of dwarf stars \citep[e.g.][]{bb90}, although some discrepancies between
the strengths of atomic and molecular absorption features have been noted
\citep{basri97}.  At cooler spectral types, more recent work has shown that
the spectra of mid- and late-M T Tauri stars deviate substantially from the
spectra of dwarf stars.  This appears to be a consequence of the gravity
sensitive molecular absorption bands \citep[e.g. TiO, VO;][]{tw93} that
dominate the spectra at these temperatures.  In these cases, reasonable
matches to their spectra, at low resolution, have been obtained by
averaging dwarf and giant spectra \citep[e.g.][]{luhman99}.

\begin{figure}
\epsscale{0.75}
\plotone{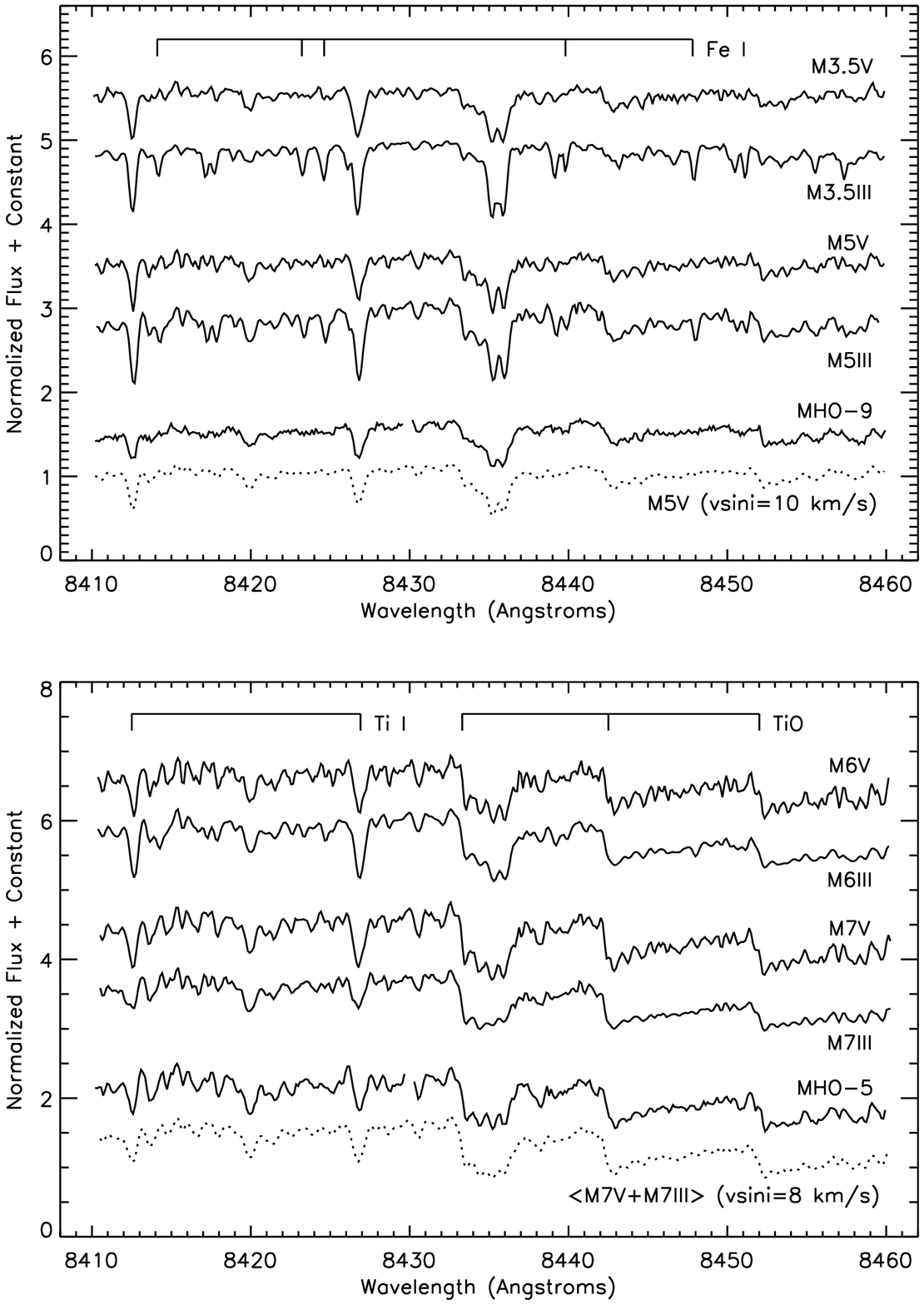}
\caption{Luminosity sensitive atomic features such as Fe I distinguish
dwarfs from giants at early- to mid-M spectral types (\textit{top panel}).
Similar to dwarf stars, these features are weak in the spectra of early- to
mid-M T Tauri stars.  The M5 T Tauri star MHO-9 is shown as an example;
it's spectrum is best matched by an M5 dwarf rotationally broadened to
$v$sin$i$ $= 10$ km/s (\textit{dotted spectrum}).  At cooler spectral
types, molecular absorption bands such as TiO and VO dominate the spectra
of both dwarf and giant stars, but the strength of these absorption bands
is luminosity sensitive (\textit{bottom panel}).  The spectra of mid- to
late-M T Tauri stars have molecular absorption bands with strengths
intermediate between that of dwarfs and giants.  The M7 T Tauri star MHO-5
is shown as an example; it's spectrum is best matched by an M7 dwarf-giant
average (\textit{dotted spectrum}).
\label{fig_viii}}
\end{figure}

With high resolution spectra, we can compare the spectra of mid- to
late-M dwarfs and giants to those of T Tauri stars in better detail.
The top panel of Figure \ref{fig_viii} shows high resolution
spectra (8410 \AA - 8460 \AA) of four early/mid-M dwarfs and giants and the 
LMTT MHO-9.  As the Figure illustrates, gravity sensitive atomic
features such as Fe I, that exhibit strong absorption in giants but weak
absorption in dwarfs, are weak in the spectra of T Tauri stars with
early- to mid-M spectral types.  Consequently, we find that the spectra of
$\sim$M5.5 and hotter T Tauri stars are best matched by the spectra of
dwarfs, as opposed to giants or some combination of the two.  The bottom
panel of Figure \ref{fig_viii} shows high resolution spectra (8410 \AA -
8460 \AA) of four mid- to- late M dwarfs and giants and the LMTT MHO-5.  At
these cooler spectral types the majority of gravity sensitive
atomic features have disappeared and the spectra are very strongly
dominated by the molecular absorption bands.  In the spectra of mid- and
late-M T Tauri stars, the strength of these absorption bands appears
intermediate in strength between that of dwarfs and giants.  Similar to
previous work at low spectral resolution, we find the combined spectra of
equally weighted dwarfs and giants provide better matches to the spectra of
$\sim$ M6 and cooler T Tauri stars than either dwarfs or giants of any
spectral type.  Thus, in determining the spectral types and modeling the
photospheric emission (\S 3.1.3), dwarf stars are used for spectral types
hotter than M6 and averages of dwarfs and giants are used for spectral
types of M6 and cooler.

With these comparison standards, the spectral types of the LMTTs
were determined primarily from the temperature sensitive region of 8360 \AA
- 8480 \AA\, (Figure \ref{fig_viii}), although the types inferred were
confirmed with other portions of the spectra.  
The relatively low rotational velocities (3.1.2) and the, at most, modest
continuum excesses at the reddest wavelengths (\S 3.1.3) have little effect
on the observed atomic-line and molecular-band ratios that distinguish
the spectral types.  Consequently, spectral types are determined with a
precision of $\pm 0.5$ spectral subclasses, even for the previously
classified continuum T Tauri stars.  We caution, however, that because of
the intrinsic differences between LMTTs and both dwarf and giant
stars, the inaccuracy of the spectral classification maybe larger (e.g. 1
spectral subclass).  The determined spectral types are listed in Table 1,
ordered from hottest to coldest.  Table 1 also lists the previous
spectral types assigned to these stars based on low resolution spectra.
The new spectral types of the coldest LMTTs are in many cases 0.5
spectral types cooler than previously inferred.  This systematic offset
most likely stems from using a different sample of comparison standards.
Comparisons with mid- and late-M dwarf stars lead to slightly warmer
spectral types than comparisons with the averages of dwarf and giant 
stars.

\subsubsection{Radial and Rotational Velocities}

The radial velocities of the observed LMTTs are determined via a
cross-correlation analysis \citep[e.g.][]{hartmann86}.  The
spectral orders are divided up into 20 \AA\, bins and those without
prominent telluric absorption features \citep[e.g.][]{tr98}, strong gravity
sensitive features (K I 7665 \& 7699 \AA, Na I 8183 \& 8195 \AA, Ca II 8662
\AA), or stellar emission-lines are cross-correlated with the M6 dwarf Gl
406.  This comparison star was observed during the same night as the T
Tauri stars.  A radial velocity of $19.175 \pm 0.110$ km/s is adopted for
Gl 406; this value is based on 10 measurements spanning 1800 days
(X. Delfosse 2001, private communication).  The positional and
time-of-night velocity corrections were determined using the
\textit{rvcorrect} task in IRAF.  Uncertainties are based on the standard
deviation of the measurements from all orders.  The radial velocities are
listed in Table 1.  The values listed for GG Tau Ba, GG Tau Bb, and UX Tau
C are from \citet{wgrs99} and \citet{bm95}.

The observed radial velocities range from 14.7 km/s to 20.8 km/s, with a
mean of $17.1$ and standard deviation of $1.8$ km/s.  These values are
consistent with the radial velocities for other Taurus T Tauri stars
\citep[$17.4$ km/s and $2.1$ km/s, respectively;][]{hartmann86}, and
supports an origin within this star forming region.  Additionally, these
consistent radial velocities imply that these LMTTs are unlikely to be
short-period ($\lesssim$ 100 days) spectroscopic binaries, but additional
epochs of spectroscopy are needed to confirm this.

The rotational velocities ($v$sin$i$) for the observed targets are
determined from the width of the peak in their cross-correlation with a
slowly rotating standard.  The width is estimated by fitting a Gaussian plus
quadratic function to the cross-correlation peak.  The width versus
$v$sin$i$ relation is determined empirically by cross-correlating a
'spun-up' spectrum of a slowly rotating standard.  The rotationally
broadened spectra are constructed using the profiles given in Gray (1992;
$\epsilon = 0.6$).  Uncertainties are based on the standard deviation of
the measurements from different orders.  
We note that although a continuum excess (see \S 3.1.3) will
diminish the effective height of the peak of the cross-correlation
function, we find that this has little effect on the width of the peak for
modest or better S/N spectra (S/N $\gtrsim$ 30).  For example, consistent
$v$sin$i$'s are derived for the 'continuum' T Tauri stars at both the
faint, more heavily veiled blue end and the brighter, less veiled red end.

The intrinsic differences between the spectra of LMTTs and slowly rotating
comparison standards (either dwarfs and giants) affect our $v$sin$i$
measurements and therefore merit some discussion.  With a velocity
resolution of 8.8 km/s and modest signal-to-noise ratios, we can measure
rotational velocities down to roughly 6 km/s ($\sim 8.8 / \sqrt{2}$).
Using the M6 dwarf Gl 406 as a comparison standard \citep[which has a
$v$sin$i$ of $< 2$ km/s;][]{delfosse98}, the $v$sin$i$'s measured for all
the LMTTs $and$ the
late-type giants are greater than 9 km/s.  Since it is unlikely that the
giants have a measurable rotation, we attribute these large $v$sin$i$
velocities in their case to the broadening of molecular features in their
spectra because of lower surface gravity (the opposite of what happens to
atomic lines in warmer giants).  The effect can be seen in the M6 and M7
standard spectra shown in Figure 1; the TiO bands of the giants are broader
and thus appear smoother than in the dwarfs.  A similar effect is seen in
the spectra of \citet{mt99}. We are not sure of the reason
for this effect.  Possible explanations include a weakening continuum and 
strengthening of molecular bands at low gravity and temperature, an
increase in the turbulent velocities in lower gravity atmospheres, or
the temperature mis-match between dwarfs and giants of the same
spectral type.  Resolving this issue is beyond the scope of this study.

The broader molecular features of giants relative dwarfs raises the 
question of whether similar conditions in the LMTTs are causing
similar line broadening compared to dwarf standards.  If so, then
the $v$sin$i$ velocities based on dwarf standards would be over-estimates.
In support of this, the use of giants as rotational standards yields many
slow $v$sin$i$ values (below our detection limit) for the LMTTs.  On the
other hand, early M and hotter TTS do not all have broader lines than
dwarfs, and
some rotations down to 5 km/s have been inferred using dwarf standards
\citep{bb90}.  Furthermore, model spectra from Allard and Hauschildt
(private communication) show no substantial difference in the linewidths at
log(g)=4 and log(g)=5 in this temperature range, values typical for T Tauri
stars and dwarfs, respectively.  This issue can only be resolved with a
more detailed atmospheric analysis and further studies of cool giants
compared with dwarfs (at high spectral resolution). The identification of
LMTTs with lines narrower than that of giants would also be definitive.

In assessing the spectral types of our LMTTs, we found it best to use a
combination of giant and dwarf spectra for the latest types.  We adopt the
same procedure for choosing a comparison standard in our rotational
analysis (Table \ref{tab_spec}).  Consequently, the $v$sin$i$ values for
the latest types are intermediate between what is found using dwarf and
giant standards, being about 3 km/s less than the dwarf values and 4-5 km/s
higher than the giant values.  All values should be considered provisional,
and subject to change with a a better understanding of the molecular line
broadening in LMTTs.  On the other hand, the conclusions we draw from these
rotational velocities (\S 4.1) are not likely to undergo a qualitative
change even after this issue is resolved.

\subsubsection{Continuum Excesses}

A continuum excess, called $r$ and defined as $r =
F_{excess}/F_{photosphere}$, is measured by comparing the depths of
photospheric features in the spectrum of a T Tauri star with the depths of
the same features in the spectrum of a standard star.  This procedure
was used to search for any continuum excesses associated with the LMTTs
studied here.  The T Tauri spectra are compared to
the best-matched spectra identified in \S 3.1.1 plus an excess
continuum that is flat over each 20 \AA\, interval for which the fits are
calculated.  The standard stars are first rotationally broadened to have
the same $v$sin$i$ as the LMTTs.  The amount of excess emission is
then determined by minimizing $\chi^2$ of the model fit, a procedure
similar to that of \citet{hartigan89, hartigan91}.  The continuum excesses
added to the standard are allowed to be negative - this results in a
comparison standard with deeper absorption features.  Consequently, stars
with no continuum excess should have $r$'s that scatter about zero, and
this scatter can be used to estimate the uncertainty in $r$.

\begin{figure}
\epsscale{1.0}
\plotone{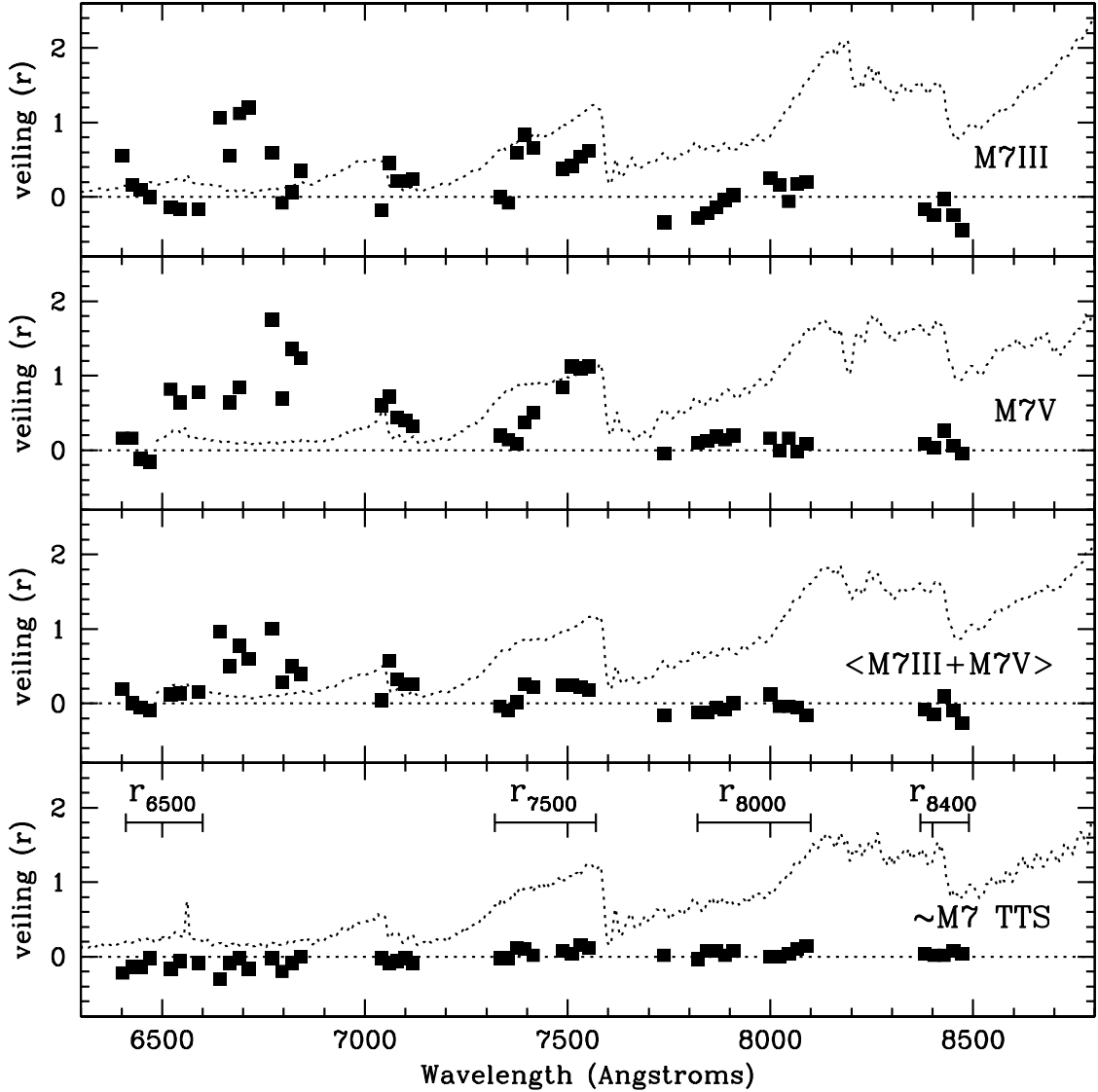}
\caption{Measurements of the MHO-5's continuum veiling versus wavelength
are determined using an M7 giant, an M7 dwarf, an M7 dwarf-giant average,
and MHO-4.  The large wavelength dependent veiling values, especially near
6700 \AA, are indicative of poor matches between MHO-5 and the template
used, and not because of a real continuum excess.  The best match is
provided by the M7 T Tauri star MHO-4, although portions of the M7
dwarf-giant average give consistent zero veiling levels.  The 4 regions
over which veiling estimates are extracted are shown.  Low resolution
spectra of the indicated spectral types are also shown (\textit{dotted
lines}), normalized at 7500 \AA\, as a reference for the underlying stellar
continuum.
\label{fig_mho5}}
\end{figure}

The uncertainties used in minimizing the $\chi^2$ fits are statistical and
do not represent the major source of error in these fits, which are the
spectral- and luminosity-class mis-matches between the LMTTs and
the available standard spectra, especially at the coolest spectral types.
As an example, shown in Figure \ref{fig_mho5} are the veiling measurements
for the M7 star MHO-5, determined using an M7 giant, an M7 dwarf, an M7
dwarf-giant average, and the M7 T Tauri star MHO-4.  Low resolution
spectra of these spectral types are also shown as a guide for where the
fits are being calculated.  The large, wavelength-variable
veilings implied at wavelengths short-ward of 7600 \AA\, using the dwarf and
giant spectra are indicative of a poor spectral match; a featureless
spectrum (i.e. highly veiled) usually provides a better fit than a spectrum
with different features.  Portions of the MHO-5 spectrum short-ward of
7600 \AA\, can be reasonably represented by the dwarf-giant average with no
veiling, but the best fit is attained using the T Tauri star MHO-4 with no
veiling.  At longer wavelengths, all comparison stars yield at least a
reasonable match to MHO-5, but again the T Tauri star MHO-4 provides the
best fit.  The optimal fit provided by MHO-4 is not surprising.  Previous
veiling studies of hotter stars have found that non-veiled T Tauri stars
generally work better as comparison standards than dwarfs
\citep[e.g.][]{hartigan91}.  Since we do not have spectra of non-veiled T
Tauri stars spanning the full range of spectral types of the sample studied
here, we use the few we do have to identify wavelength
regions for which T Tauri spectra can be best matched by dwarf or
dwarf-giant averages.  First, we conclude that the 3 approximately M7 T Tauri
stars MHO-4, MHO-5, \& V410 Tau X-ray 3 are non-veiled stars based on (1) the
zero veiling implied from fits against each other, (2) the zero veiling
implied over the majority of the spectral coverage by using the M7
dwarf-giant average, and (3) the narrow H$\alpha$ emission profiles of
these stars (\S 3.1.4).  
Four regions are identified where dwarf-giant averages (or dwarfs at cooler
spectral types) provide a consistent match to the T Tauri spectra with no
veiling.  These veiling regions are $r_{6500}$ (6415 - 6595 \AA),
$r_{7500}$ (7325 - 7565 \AA), $r_{8000}$ (7815 - 8095 \AA), and $r_{8400}$
(8370 - 8490 \AA), and are shown in Figure \ref{fig_mho5}.

Veiling estimates are derived for each T Tauri star within these 4
wavelength regions, in 20\AA\, intervals, as described above.  The veiling
is considered a detection if the mean veiling within each region is larger
than one standard deviation.  Otherwise an
upper limit to the veiling is set at one standard deviation.  The continuum
excess is assumed to be constant over each of these wavelength regions
(120 \AA $-$ 280 \AA\, in length).  However, a straight mean of the veiling
estimates within each veiling region is not an accurate measurement of this
excess.  Because of the strongly wavelength dependent molecular absorption
bands, the underlying continuum, and thus the ratio of F$_{excess}$ to
F$_{photosphere}$, can vary dramatically within each region.  Therefore,
the veiling estimates of each 20 \AA\, interval are scaled by the ratio of
the underlying photospheric flux to the photospheric flux at a normalizing
wavelength (ie. 6500 \AA, 7500 \AA, 8000 \AA, and 8400 \AA).  Low resolution
dwarf spectra of the same spectral-type are used to determine the
photospheric normalization.  The veiling within each veiling interval is
then estimated from the mean of these normalized veiling measurements and
the uncertainties are estimated from their standard deviation.  These
values are listed in Table \ref{tab_excess}.

\subsubsection{Prominent Emission Lines}

The strongest emission line in the observed spectra is
H$\alpha$.  The EW[H$\alpha$] values are listed in Table \ref{tab_spec}
and are, with the exception of V410 Tau X-ray 3, all systematically less
than the EW[H$\alpha$]s previously measured from low resolution
spectroscopy (see Table \ref{tab_spec} for references to previous
spectra).  This highlights an important, but often unrecognized point.  The
H$\alpha$ feature resides near the 6569 \AA\, TiO band-head, a strong
feature in mid- and late-M stars \citep{kirkpatrick91, 
tr98}.  In low resolution spectra, the edge of this band-head becomes blended
with the H$\alpha$ emission feature, leading to an under-estimate of the
continuum on the red-ward side and thus and an over-estimate of the
EW[H$\alpha$].  Consequently, low resolution spectra systematically
over-estimate the EW[H$\alpha$] for mid-M and cooler stars, an effect that
can lead to the mis-classification of T Tauri types (\S 4.2).

In Figure \ref{fig_hatts} are shown the H$\alpha$ velocity profiles of the
LMTTs and the M6 dwarf Gl 406.  For the 7 LMTTs with no detected veiling,
the H$\alpha$ profiles are relatively symmetric and narrow, similar to
those of weak-lined T Tauri stars (WTTSs), which
are not thought to be accreting.  Full widths at 10\% of the peak
flux levels (called 10\%-widths, hereafter) for these stars are all less
than 200 km/s.  In contrast, the 3 veiled LMTTs (GN Tau, CIDA-1, and GM
Tau) exhibit broad H$\alpha$ profiles similar to classical T Tauri stars
(CTTSs; \S 4.2).  The 10\%-widths for these 3 stars are 525 km/s, 375 km/s,
and 370 km/s, respectively, and are consistent with the predictions of
magnetically channeled accretion flow \citep[e.g.][]{hartmann94}.  In
addition, the veiled stars show blue-shifted absorption components
superimposed upon the H$\alpha$ emission profiles.  This absorption is
common in CTTSs and usually attributed to a strong wind
\citep[e.g.][]{edwards87, ab00}.

The 3 veiled LMTTs also exhibit, within the wavelength coverage of our
spectra, an array of other strong emission lines such as Fe II (6456 \AA),
He I (6678 \AA, 7066 \AA), O I (8446 \AA), and Ca II (8662 \AA).  Most of
these emission features are only seen in very high accretion CTTSs
\citep[e.g. DG Tau, DR Tau;][]{hg97}. 

\subsubsection{Na I Absorption}

Figure \ref{fig_na} shows the gravity sensitive Na I doublet at 8183 \AA\,
and 8195 \AA, for an M6 giant (HD 78712), an M6 dwarf (Gl 406), and 3 mid-M
LMTTs.  The wing strengths of the LMTTs are intermediate between
that of the narrow giant and strongly pressure broadened dwarf.  A similar
effect is seen with the K I features at 7665 \AA\, and 7699 \AA, although
these lines are badly blended with the telluric A-band.  The LMTTs appear
to have moderately lower gravities than main-sequence stars, consistent
with their pre-main sequence classification.

\subsection{Mass and Age Estimates}

The standard method for estimating masses and ages for young stars is to
compare their observed luminosities and temperatures with the predictions
of theoretical pre-main-sequence (PMS) evolutionary models.  In our
analysis, we adopt the PMS evolutionary models of \citet{bcah98} that are
(1) calculated with more realistic non-grey atmospheres, (2) predict
stellar masses that are reasonably consistent with dynamically inferred
values \citep{simon00}, and (3) predict coeval ages for the young quadruple
GG Tau \citep{wgrs99} and young cluster IC 348 \citep{luhman99}, using a
temperature scale intermediate between that of dwarfs and giants.  A common
difficulty in positioning low mass young
stars in the H-R diagram is that this requires knowledge of their intrinsic
colors and bolometric corrections from which the extinctions and
luminosities are estimated.  This is especially problematic for stars with
large continuum excesses that may bias the inferred luminosities.
Fortunately, mass estimates are relatively insensitive to uncertainties in
luminosity since young ($< 10$ Myrs) low mass stars ($< 0.5$ M$_\odot$) lie
on the convective and primarily vertical part of PMS evolutionary tracks.
At these masses and ages, mass estimates are predominately determined by
their effective temperatures.

\citet{wg01} have shown that the young stars in Taurus have a mean age of
2.8 Myrs using the \citet{bcah98} evolutionary models at sub-solar masses.
Their study, which includes many of the young stars studied here, shows
that most stars in Taurus-Auriga are coeval to within a few million years
and that there is no evidence for a mass dependence on the stellar age.
Therefore, we adopt an age of 3 Myrs for the LMTTs sample.  Masses
are then estimated by first converting the spectral types determined in \S
3.1.1 to effective temperatures using the intermediate temperature scale of
\citet{luhman99}.  These effective temperatures are then compared with the
predictions of the 3 Myr isochrone of \citet{bcah98} to estimate masses.
Uncertainties in the masses are estimated from the uncertainties in the
spectral types (0.5 spectral sub-classes).   This methodology allows us to
estimate consistent masses for the 3 continuum T Tauri stars, which have no
measured optical colors, and for the unresolved binary GN Tau.  The
inferred masses are listed in Table \ref{tab_prop}.  We note that masses
determined directly from luminosity and temperature estimates (ie. with no
age assumption) for the subsample of stars with optical measurements, and
masses determined assuming different ages (1 Myrs, 5 Myrs) are all
consistent with the values listed in Table \ref{tab_prop} to within the
uncertainties; as emphasized above, the masses are primarily determined by
the adopted temperatures.

At an assumed age of 3 Myrs, the inferred masses range from $0.46 \pm
0.09$ M$_\odot$ down to $0.042 \pm 0.015$ M$_\odot$.  Although the adopted
methodology yields reasonably precise mass estimates, we caution that the
absolute uncertainties in these values could be as large as 50\% because of
uncertainties in the evolutionary models and the temperature scale.
Nevertheless, under the adopted assumptions, five objects have
masses below the hydrogen-burning minimum mass of $\sim 0.075$ M$_\odot$
\citep{cb97}, and thus are T Tauri brown dwarfs.

\subsection{Mass Accretion Rates}

The ultra-violet and optical continuum excess emission (or veiling) often
observed in the spectra of T Tauri stars is attributed to the high
temperature regions generated as circumstellar material accretes from the
disk onto the star.  Measurements of this excess can therefore be used to 
estimate the mass accretion rate.  To accomplish this, we use the continuum
excesses (or upper limits) measured at 6500 \AA, which are the closest in
wavelength to where the continuum excess dominates.  First, the excess
measured at 6500 \AA\, is assumed to be equivalent to the continuum excess
within the $R_c$ broad-band filter.  The $R_c$ magnitudes of the continuum
excess are then calculated from the observed $r_{6500}$ values and the
$R_c$ photospheric magnitudes as predicted by the adopted evolutionary
models \citep[see \S 3.2;][]{bcah98}.  The total accretion luminosity can
be estimated from these $R_c$ excess magnitudes with a bolometric
correction ($M_{BOL} = M_{R_C} + BC$).  We adopt a bolometric correction
of -0.4 mag, which is characteristic of a marginally optically thick slab
or pure hydrogen gas at $10^4$ K and $n = 10^{14}$ cm$^{-3}$
\citep{heg95}.  Although more sophisticated models for the accretion
luminosity have been proposed, it is not clear that they can successfully
reproduce the continuum excess at red optical wavelengths better than
single temperature slab models \citep[see e.g.][]{cg98}.  We prefer the
single temperature model since it is similar to models used in previously
studies \citep{valenti93, heg95, gullbring98} and therefore allows us to
make more
consistent comparisons with these measurements.  These accretion
luminosities are then converted to mass accretion rates following the
simple infall model of \citet{gullbring98}, in combination with radius and
mass estimates inferred from the evolutionary models (\S 3.2).  These
values are listed in Table \ref{tab_prop}.  Mass accretion rate upper
limits are inferred for stars with veiling upper limits.  
Based on uncertainties in the measured veiling estimates and unknown
inclinations of each system, these mass accretion estimates should be
accurate, on a relative scale, to within factors of 2-3.  More fundamental
uncertainties in the accretion model (e.g. bolometric correction, inner
disk radius) imply that the mass accretion rates estimates may only be
accurate, on an absolute scale, to an order of magnitude.

GN Tau, CIDA-1 and GM Tau have mass accretion rates of $1.3 \times 10^{-8}$ 
M$_\odot$yr$^{-1}$, $3.1 \times 10^{-9}$ M$_\odot$yr$^{-1}$, and
$2.4 \times 10^{-9}$ M$_\odot$yr$^{-1}$, which are typical for $\sim 0.5$
M$_\odot$ T Tauri stars \citep[e.g.][]{gullbring98}.  However, both CIDA-1
and GM Tau have masses several times below this typical value (0.12
M$_\odot$ and 0.073 M$_\odot$, respectively).  GM Tau is of particular
interest.  With a mass just below the stellar/substellar boundary, it is
the only known brown dwarf with unambiguous accretion signatures.  The mass
accretion rate upper limits for the remaining sample range from $3.3 \times
10^{-10}$ M$_\odot$yr$^{-1}$ to $3.2 \times 10^{-11}$ M$_\odot$yr$^{-1}$
for a sample of masses spanning 0.17 M$_\odot$ to 0.043 M$_\odot$.

We note that a veiling upper limit of 0.1 in the $R_c$ band (about the best
that is achievable using comparison dwarf-giant spectra), corresponds to a
mass accretion rate upper limit of $\sim 10^{-10}$ M$_\odot$/yr for a 0.1
M$_\odot$ star, and is therefore the approximate mass accretion rate
detection limit of the adopted model.  Smaller mass accretion rates may be
detectable by observing at shorter wavelengths or by modeling the profiles
of permitted emission lines.  As an example of the latter,
\citet{muzerolle00a} estimate a mass accretion rate of $\sim5 \times
10^{-12}$ M$_\odot$yr$^{-1}$ for the low mass T Tauri star V410 Anon 13
from model fitting its H$\alpha$ profile; this is the smallest mass
accretion rate yet determined for a T Tauri star.  But even at this low
level of mass accretion, the H$\alpha$ 10\%-width of V410 Anon 13 is $\sim$
250 km/s \citep{muzerolle00a} and is distinctly larger than the 10\%-widths
for the stars with no measurable veilings.  This suggests that if our
non-veiled LMTTs are experiencing any accretion, it is probably at a rate
less than $5 \times 10^{-12}$ M$_\odot$/yr.

\section{Discussion}

The signatures of youth, location, and radial velocities of the LMTTs
studied here compellingly support the claim that
they are members of the Taurus-Auriga star forming region.  We use their
properties to investigate the rotation rates, the mass accretion rates,
signatures of circumstellar accretion, and possible formation scenarios for
LMTTs in Taurus-Auriga.

\subsection{Rotational Properties}

The accretion of high angular momentum material during the earliest stages
of star formation is expected to produce very rapidly rotating stars
\citep[e.g.][]{durisen89}, with rotational velocities comparable to the
break-up velocity ($v_{br} = \sqrt{GM/R} \sim 300$ km/s).  In contrast
to this prediction, young stars in Taurus rotate slowly, with rotational
velocities that are typically less than one-tenth $v_{br}$
\citep[e.g.][]{hartmann86, bouvier90, bouvier95}.  Proposed explanations
for the slow rotation rates usually involve ``magnetic braking'', a
mechanism which involves the star being magnetically coupled to the
accretion disk and transferring angular momentum to the slower rotating
outer parts of the disk, or possibly a stellar wind \citep{konigl91,
shu94}.  The slow rotation of young stars in Taurus, however, is not common
to all star forming regions.  Photometric monitoring studies of stars in
Orion show that a large fraction (30\%) have rotational velocities that are
larger than 0.2 times their rotational break-up velocity \citep{stassun99}.
\citet{cb00} present a critical comparison of the rotational velocity
distributions of Taurus and Orion and show that they are
different at the $>$3$\sigma$ level.  However, one
significant caveat in the comparison of Taurus and Orion rotational
velocities is that the Taurus sample is biased towards higher masses (87\%
$>$ 0.4 M$_\odot$ for Taurus versus 40\% $>$ 0.4 M$_\odot$ for Orion).
In Orion it is predominantly the lowest mass stars (M $<$ 0.2 M$_\odot$)
that are the rapid rotators \citep{herbst01}.  With this in mind, we use
the mass and $v$sin$i$ estimates derived here to investigate the mass
dependence of rotation in Taurus, and to establish a less biased comparison
between the rotational distributions of Taurus and Orion.

Following \citet{cb00}, the observed $v$sin$i$ values are normalized by
their break-up velocity; mass and radius estimates are obtained
from the adopted evolutionary models (\S
3.2).  These values are plotted in Figure \ref{rot_mass}, along with the
sample of higher mass\footnote{\citet{cb00} calculate masses using use the
evolutionary models of \citet{siess00}.} Taurus stars studied by
\citet{cb00}.  None of the low mass stars and brown dwarfs studied here
rotate with speeds of more than 20\% of their break-up velocity, similar to
the distribution of higher mass stars in Taurus.  A K-S test shows that the
distributions of $v$sin$i$/$v_{br}$ for stars above and below 0.2 M$_\odot$
are indistinguishable.  Even at masses which extend into the substellar
regime, Taurus does not appear to produce rapidly rotating objects.  This
result contrasts significantly with the large fraction (30\%) of stars in
Orion which rotate faster than 20\% of their break-up velocity.  Thus,
in agreement with \citet{cb00}, we conclude that Taurus and Orion produce
different rotational distributions.

The variations in the rotational distributions of star forming regions
pose an interesting problem for theories of low mass star formation.  Two
proposed explanations for the more rapidly rotating Orion population,
relative to Taurus, are (1) weaker magnetic fields, and thus weaker
magnetic brakes, for stars in Orion relative to Taurus \citep{cb00}, and
(2) younger ages and thus less effective braking for stars in Orion
relative to Taurus \citep{hartmann02}.  It should be realized, however,
that there is at most only marginal evidence for a difference in the
magnetic field strengths \citep{gcs95, garmire00} or ages
\citep[cf.][]{luhman00a, luhman00b} of stars in Orion and Taurus.
Perhaps more fundamentally, several rotational studies have challenged
the role of disk braking at early evolutionary stages \citep{stassun99,
stassun01, rebull01}. \citet{stassun01}, for example, suggest initial
conditions may be more important than disk braking in determining the
rotational velocities at T Tauri ages.  In support of this, we note that
the mass dependence of the rotational velocities in Orion are not easily
explainable through disk braking evolution alone.  Following
\citet{hartmann02}, the disk braking timescale scales as
$M_\star\dot{M}^{-1}f$, where $M$ is the mass of the star, $\dot{M}$ is the
mass accretion rate, and $f$ is a measure of the rotational angular
velocity normalized by the breakup angular velocity.  Thus, if the mass
accretion rate, $\dot{M}$, is roughly proportional to $M$ as suggested in
\S 3.3, then the disk braking timescale should be independent of mass
unless $f$, the \textit{initial} angular momentum distribution, is mass
dependent.  This argues that the rapid rotation of low mass Orion members
is not a consequence of evolution, but a result of the conditions set up in
the formation process.  The difference between rotational distributions of
Taurus and Orion may also be a result of different initial conditions, and
possibly related to Taurus's high binary fraction \citep{gws02}, low
stellar density \citep{jh79}, and less turbulent cloud cores
\citep{myers98}.

Finally, we note that the LMTTs in Taurus (M $<$ 0.2 M$_\odot$) have radii,
determined from both pre-main sequence models and luminosity estimates,
that are 4-6 times their main sequence values.  These radii are roughly
twice that of more massive T Tauri stars \citep[e.g.][]{bouvier95}.
Under the assumption that disk braking terminates simultaneously for high
and low mass T Tauri stars, continued contraction toward the main sequence
will transform the mass independent rotational distribution into a mass
dependent distribution, with the lowest mass stars and brown dwarf rotating
the fastest.  If the diminished frequency of accreting disks among LMTTs
implies disk braking terminates more quickly at low masses, then the
resulting zero-age main sequence rotational distribution may exhibit
and even stronger mass dependence.  These effects, in part, may lead to the
observed rapidly rotating population of very low mass stars and brown
dwarfs in the field \citep{basri00}.

\subsection{Distinguishing CTTSs and WTTSs based on H$\alpha$ emission}

The most common criterion used to distinguish CTTSs from WTTSs is H$\alpha$
emission.  Although WTTSs usually exhibit some H$\alpha$ emission, which is
attributed to their active chromospheres, this emission is limited in
amount by the saturation level of the chromosphere, and is limited in
profile-width by the stellar rotation and mirco-turbulent, non-thermal
velocities of the chromosphere \citep[e.g.][]{jds00}.  In contrast to this,
CTTSs usually show strong, broad H$\alpha$ emission profiles generated from 
the high temperatures and high velocities of accreting circumstellar
material \citep[e.g.][]{hartmann94}.  These properties are illustrated in 
Figure \ref{width}, which plots the EW[H$\alpha$]s versus the 10\%-widths
of H$\alpha$ for a large sample of T Tauri stars.  The measurements were
extracted from high resolution spectra presented in \citet{heg95},
\citet{ab00}, \citet{muzerolle00b}, and \S 3.1.4.  Multiple measurements for
a star are averaged and double-lined spectroscopic binaries are excluded (DQ
Tau, V826 Tau, Hen 3-600A, TWA 5A, TWA 6).  From this sample, stars that
exhibit an optically veiled spectrum ($r > 0.06$) are classified as CTTSs,
while stars with no optical veiling are classified as WTTSs.  Using the
presence of optical veiling to identify accretion, we develop empirical
criteria for distinguishing CTTSs and WTTSs.

Historically, the EW[H$\alpha$] has been used to distinguish between WTTSs
and CTTSs \citep[e.g.][]{hb88}.  As shown in Figure \ref{width}, however,
no unique value of EW[H$\alpha$] distinguishes all CTTSs from WTTSs.  This
is primarily a consequence of the "contrast effect" \citep[cf.][]{bm95}.
For example, H$\alpha$ emission from equally
saturated chromospheres of a late-M star and an early K-star will appear
much more prominently in the M star because of its substantially diminished
photospheric continuum near 6500 \AA.  \citet{martin98} suggests
EW[H$\alpha$] criteria that account for this spectral type dependence.  We
suggest a slight modification to these values based on the large sample of
T Tauri stars presented in Figure \ref{width} that extend to cooler
spectral types than previously considered.  Specifically, we propose that a
T Tauri star is classical if EW[H$\alpha$] $\ge$ 3 \AA\, for K0-K5 stars,
EW[H$\alpha$] $\ge$ 10 \AA\, for K7-M2.5 stars, 
EW[H$\alpha$] $\ge$ 20 \AA\, for M3-M5.5 stars, and
EW[H$\alpha$] $\ge$ 40 \AA\, for M6-M7.5 stars.
These values are determined empirically from the maximum EW[H$\alpha$]s for
non-veiled T Tauri stars within each spectral type range (see Figure
\ref{width}).  Stars with EW[H$\alpha$] below these levels are not
necessarily WTTSs, however.  Confirmation depends upon the Li abundance
\citep[cf.][]{martin98}.  Assuming that optical veiling correctly
identifies accretion, the EW[H$\alpha$] classification is correct 95\% of
the time.  We caution that these values were determined from high spectral
resolution measurements (R $>$ 20,000) and the biases in EW measurements
using lower resolution spectra, as noted above, may result in slightly
different (larger) distinguishing values.

As Figure \ref{width} demonstrates, optically veiled stars are also
distinguishable from stars with no optical veiling based on the 10\%-width
of H$\alpha$ emission.  In all but one case, stars with 10\%-widths greater
than $270$ km/s are optically veiled, while stars with narrower 10\%-widths
are not.  The one exception is UX Tau A, a non-veiled early-K T Tauri star
with a 10\%-width of 475 km/s.  However, extracting optical veiling
measurements for early-K T Tauri stars is difficult because of both their
increasing relative stellar luminosity and the difficulty in determining
spectral types \citep[the spectral type of UX
Tau A ranges from K0 to K5;][]{bb90, hss94}.  Thus, UX Tau A may in fact
have some low level optical veiling, as measurements by \citet{bb90}
suggest, and perhaps should be considered a CTTS.  Therefore we propose a
new accretion diagnostic: T Tauri stars with 10\%-widths $> 270$ km/s are
CTTSs.  Our choice of 10\% of the peak flux is somewhat subjective, but
this level is typically low enough to avoid biases introduced by
superimposed blue-shifted absorption features and high enough to
distinguish above the often uncertain continuum level in late-type stars.
Based on the optical veiling as an accretion diagnostic, this
classification is correct 98\% of the time (100\% if UX Tau A is
accreting).  10\%-widths measurements are advantageous to optical veiling
measurements because they (1) can be extracted over a short wavelength
range, (2) do not depend on the underlying stellar luminosity, and (3) do
not depend on having a properly identified comparison template.  Thus,
given the strong correlation between the presence of optical veiling
and broad 10\%-widths, H$\alpha$ 10\%-widths may be a more accurate
diagnostic of accretion, or CTTS type, than either optical veiling or
EW[H$\alpha$].

\subsection{The Mass Dependence of Circumstellar Accretion}

In Figure \ref{mass_acc} the mass accretion rates or mass accretion rate
upper limits are plotted versus evolutionary-model masses for the 10 stars
and brown dwarfs studied here, and for the low mass T Tauri star V410 Anon
13 \citep{muzerolle00a}.  For comparison, we also plot the sample of more
massive classical T Tauri stars in Taurus from \citet{wg01}, with mass
accretion rates determined from U-band excesses following the same
accretion model that is used here \citep[i.e.][]{gullbring98} and with
stellar masses estimated from the same evolutionary model that is used
here.

\citet{wg01} and \citet{muzerolle00a} have shown evidence that the mass
accretion rate in Taurus decreases toward lower masses.  The very low
mass accretion rates of the lowest mass stars in these studies hinted at a
strong functional form ($\sim \dot{M} \propto M^{3}$).  The new accretion
rates presented here, however, demonstrate that stars at the
stellar/substellar boundary can have accretion rates comparable to higher,
more canonical mass T Tauri stars ($\sim$ K7-M0 spectral type; M $\sim$ 0.7
M$_\odot$).  Thus although the average mass accretion rate appears to
decrease with decreasing mass, it is a weaker dependence than initially
presumed.  A mass dependence of the form $\dot{M} \propto M$
(i.e. $\dot{M}/M$ is constant) is consistent with the available data
(Figure \ref{mass_acc}), but there are yet too few data points to conduct
a meaningful best fit.  We note this because circumstellar disk models with
this mass dependence match the observed NIR emission from young low mass
stars and brown dwarfs \citep{nt01}.  

\citet{rebull00, rebull02} also find evidence of a similar mass dependence
over the mass range 0.25 - 1 M$_\odot$, based on ultra-violet excesses of
young stars in Orion Nebula cluster flanking fields and NGC 2264.  We
caution, however, that the Taurus sample, as well as more complete
photometric surveys \citep[e.g.][]{rebull00, rebull02}, may still suffer
selection biases caused by high accretion stars being heavily extincted and
optically fainter \citep[see][]{wg01}. A correlation between extinction and
accretion could put high accretion low mass stars below detection limits,
but leave higher mass high accretion stars still observable, yielding an
apparent mass dependence on the average accretion rate.  Although this
possibility is worth exploring further in future studies, the current
emerging trend suggests at least a modest mass dependence on the mass
accretion rate.

The difficulty in determining the mass dependence of the mass accretion rate
stems from the selection biases in the known samples of LMTTs, as well as
the paucity of \textit{accreting} LMTTs.  To demonstrate the latter, we
determine the fraction of CTTSs in Taurus with spectral types hotter than
M5, and the fraction with spectral types of M5 and cooler (for Taurus,
M5 $\sim$ 0.2 M$_\odot$).  CTTSs are distinguished from WTTSs based on
H$\alpha$ 10\%-widths, if that information is available, or the new
EW[H$\alpha$] criterion presented in \S 4.2.  Using the sample of single T
Tauri stars listed in \citet{wg01} with spectral types hotter than M5,
63\% are classical T Tauri stars (30 of 48 stars).  We note that this
is consistent with the "JHKL excess fraction" of 69\% in Taurus \citep[for
$m_K \le 9.5$;][]{haisch00}; roughly two-thirds of the higher mass stars in
Taurus appear to be classical T Tauri stars.  In contrast to this, of the
20 T Tauri stars in Taurus of spectral type M5 and cooler (\S 2) and with no
known companions \citep{gws02}, only 30\% are classical (6 of 20).  This
fraction may be even lower since half of those determined to be classical
are determined from EW[H$\alpha$]s measured from low resolution spectra,
which biases the EWs towards larger values (see \S 4.2).  We conclude that
both the average mass accretion rate and the frequency of accreting
circumstellar disks decreases toward low masses.

\subsection{Low Mass Star and Brown Dwarf Formation}

The processes involved in the formation of low mass stellar and substellar
objects are not well understood.  The isolated gravitational collapse of a
dense molecular cloud core that successfully explains solar mass star
formation may work successfully well into the brown dwarf (and possibly
planetary mass) regime (e.g. Boss 1993).  Alternatively, if the dominant
method of formation occurs within multiple systems, as appears to be the
case within the Taurus molecular cloud \citep{wg01}, dynamical interactions
could eject one or more of the lowest mass hydrostatic cores immediately
after formation.  The ejected core, removed from the reservoir of material,
would grow from its low possibly substellar mass \citep{rc01}.  Thus the
formation of brown dwarfs may simply be a consequence of early ejection.
In support of this, there is tentative evidence for an anti-correlation
between the density of stars and the density of brown dwarfs in Taurus
\citep{martin01}.

If the lowest mass objects in Taurus result from low mass cores being 
ejected from dynamical encounters, they may have a larger radial 
velocity dispersion than the $\sim$ 2 km/s observed for higher mass stars 
\citep{hartmann86} and from $^{12}$CO gas measurements \citep{ut87}.  For
example, numerical simulations of triple encounters suggest typical
ejection velocities of $\sim 4$ km/s, with occasional high velocity
ejections \citep[10\% with $\sim 10$ km/s;][]{sd95}.  This trend is not
observed for the small sample studied here.  T Tauri stars of spectral type
M5 or cooler (9 objects; M $<$ 0.17 M$_\odot$) have a velocity dispersion 
of $1.9 \pm 0.5$ km/s.  Similar velocity dispersions are determined 
for the sub-samples that are not within binary systems (6 objects; 
$\sigma_{vel} = 2.1 \pm 0.6$), or for the single stars of spectral type 
M6 or cooler (4 objects; M $<$ 0.09 M$_\odot$ $\sigma_{vel} = 2.3 \pm 
0.8$).  The decreasing frequency of circumstellar disks for lower mass 
stars and brown dwarfs is consistent, however, with an ejection scenario 
that would likely strip a young core of its associated material.
But, the slow rotation of these low mass targets suggests, at least 
indirectly, that they retained disks for a sufficient time to allow 
some disk braking to occur.  We also note that several of the lowest
mass members of Taurus are in binary systems of separations $> 10$ AU.
These relatively wide systems are also inconsistent with an ejection
scenario, which would preferentially eject single stars.  A more complete
survey is needed to address this issue properly, however
\citep[e.g.][]{gws02}.  Nevertheless, based on the radial velocity
dispersion, evidence (both direct and indirect) for circumstellar disks,
and the known binary LMTTs in Taurus, we conclude these objects have
not been ejected at high velocities in multiple star encounters.  Their
properties suggest they experienced an early evolution that is very similar
to higher-mass T Tauri stars, but with less massive disks.

\section{Summary}

We present high resolution optical spectra obtained with the HIRES
spectrograph on the W. M. Keck I telescope of seven LMTTs in
Taurus-Auriga.  These observations are combined with previous HIRES
observations of the LMTTs GG Tau Ba (M6), GG Tau Bb (M7.5), and UX Tau C
(M5) for a sample of 10 LMTTs.  The observed Li I 6708 \AA\, absorption,
low surface gravity signatures, and radial velocities confirm that all are
members of the Taurus star forming region.  No new spectroscopic binaries
are identified.  The LMTTs have spectral types ranging from M2.5 to M7.5
and masses ranging from 0.46 M$_\odot$ to 0.042 M$_\odot$.  Five of the 10
LMTTs studied appear to be T Tauri brown dwarfs.  Of particular interest is
the previously classified 'continuum T Tauri star' GM Tau, which has a
spectral type of M6.5 and a possible substellar mass of $0.073 \pm 0.013$
M$_\odot$.

The spectroscopic properties of this sample are used to understand the
formation and early evolution of objects in Taurus-Auriga with masses near
and below the stellar/substellar boundary.  None of the LMTTs in Taurus are
rapidly rotating (vsin$i$ $<$ 30
km/s).  These slowly rotating stars nevertheless exhibit prominent
chromospheric emission.  EW[H$\alpha$] of up to 40 \AA\, are observed for
non-accreting M6 to M7.5 T Tauri stars.  Cooler spectral types may show
even stronger emission levels as the photospheric continuum continues to
decrease \citep[e.g.][]{martin01}.  The slow rotation rates of very low
mass stars and brown dwarfs in Taurus are similar to the slow rotation
rates of higher mass stars in Taurus.  Taurus appears to primarily produce
relatively slow rotating stars, independent of mass.  This result
contrasts markedly with the rapidly rotating low mass population in Orion
\citep{stassun99}, and confirms the suggestion by \citet{cb00} that
Taurus and Orion produce different rotational distributions.  The mass
dependent rotational distribution in Orion is not easily explainable via a
magnetic braking dominated evolution.  Instead, variations in the initial
angular momentum distribution seem more plausible.  Such variations may
also account for the differences between Taurus and Orion.  Although the
LMTTs in Taurus currently have rotational velocities consistent with their
more massive neighbors, their larger radii (by roughly a factor of 2) and
possibly quicker disk dissipation timescales suggests that they will
produce a more rapidly rotating low mass population at main sequence ages.
This scenario is consistent with the rapid rotation observed among the very
low mass stars and brown dwarfs in the field \citep{basri00}.

We demonstrate that the H$\alpha$ full-width at 10\% of the peak is a
good diagnostic of accretion, and is possibly more accurate than either the
equivalent width of H$\alpha$ or optical veiling.  10\%-widths $> 270$ km/s
are classical T Tauri stars (i.e. accreting), independent of stellar
spectral type.

Although the LMTTs can have accretion rates comparable to that of more
massive T Tauri stars (K7-M0 spectral type), the average mass accretion
rate appears to decrease with decreasing mass.  A functional form of
$\dot{M} \propto M$ is consistent with the available data, but the
dependence is difficult to establish accurately because of the decreasing
frequency of accreting stars at low masses, and possible selection biases.
Using newly established criteria to identify circumstellar accretion, we
find that the frequency of circumstellar accretion diminishes with decreasing
mass - 63\% of stars more massive than 0.2 M$_\odot$ appear to be accreting
while only 30\% of stars less massive show signs of accretion.  The
diminished frequency of accreting disks for LMTTs, despite their lower mass
accretion rates, implies that they are formed with less massive disks than
higher-mass T Tauri stars.

The radial velocities, circumstellar properties, and known binary LMTTs do
not corroborate suggestions that many of the lowest mass members of Taurus
have been ejected from higher stellar density regions within the cloud.
Instead, these objects appear to have formed and are evolving in the same
way as solar-mass T Tauri stars, but with less massive disks and shorter
disk lifetimes.

\acknowledgements

We thank P. Hartigan and J. Muzerolle for providing H$\alpha$ profile data,
X. Delfosse for providing radial velocities of standard stars, and
L. Hillenbrand and S. Mohanty for helpful discussions.

\appendix

\section{Comparison Standards}

Several standard stars were observed as part of this program.  The spectra
of these stars and others are used in the analysis of the late-type T Tauri
spectra.  All were observed with the HIRES spectrograph on the Keck I
telescope, except for Gl 15A which was observed using the Hamilton Echelle
Spectrograph at Lick Observatory.  Stars observed with different
instruments or resolutions were resampled and/or binned to match the
resolution of the new HIRES observations presented here.  Table 4
summarizes the properties of the comparison stars used, including the
adopted spectral types, the radial and rotational velocities, and the
references for these measurements (some derived here).

\clearpage

\begin{figure}
\epsscale{1.0}
\plotone{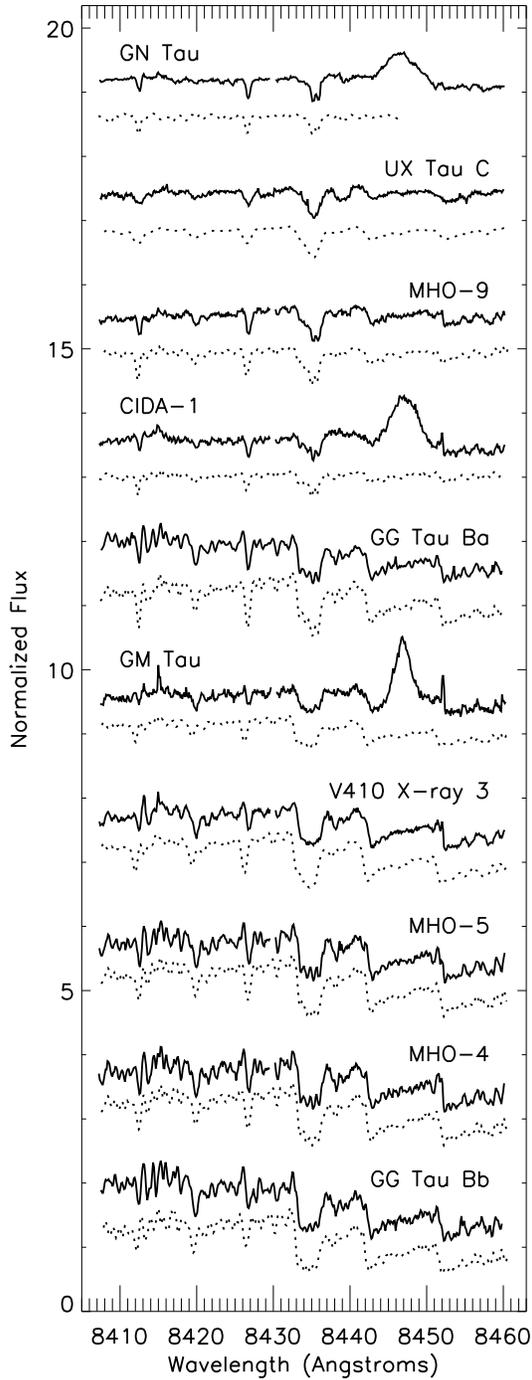}
\caption{Spectra of T Tauri stars (\textit{solid lines}) and the best fit
rotationally broadened and veiled comparison standards (\textit{dotted
lines}) from Table \ref{tab_excess}.  The stars are ordered from hottest to
coldest.  The 3 veiled stars GN Tau, CIDA-1, and GM Tau, all exhibit a
broad OI 8446 \AA\, emission.
\label{sptp_fits}}
\end{figure}

\begin{figure}
\epsscale{0.80}
\plotone{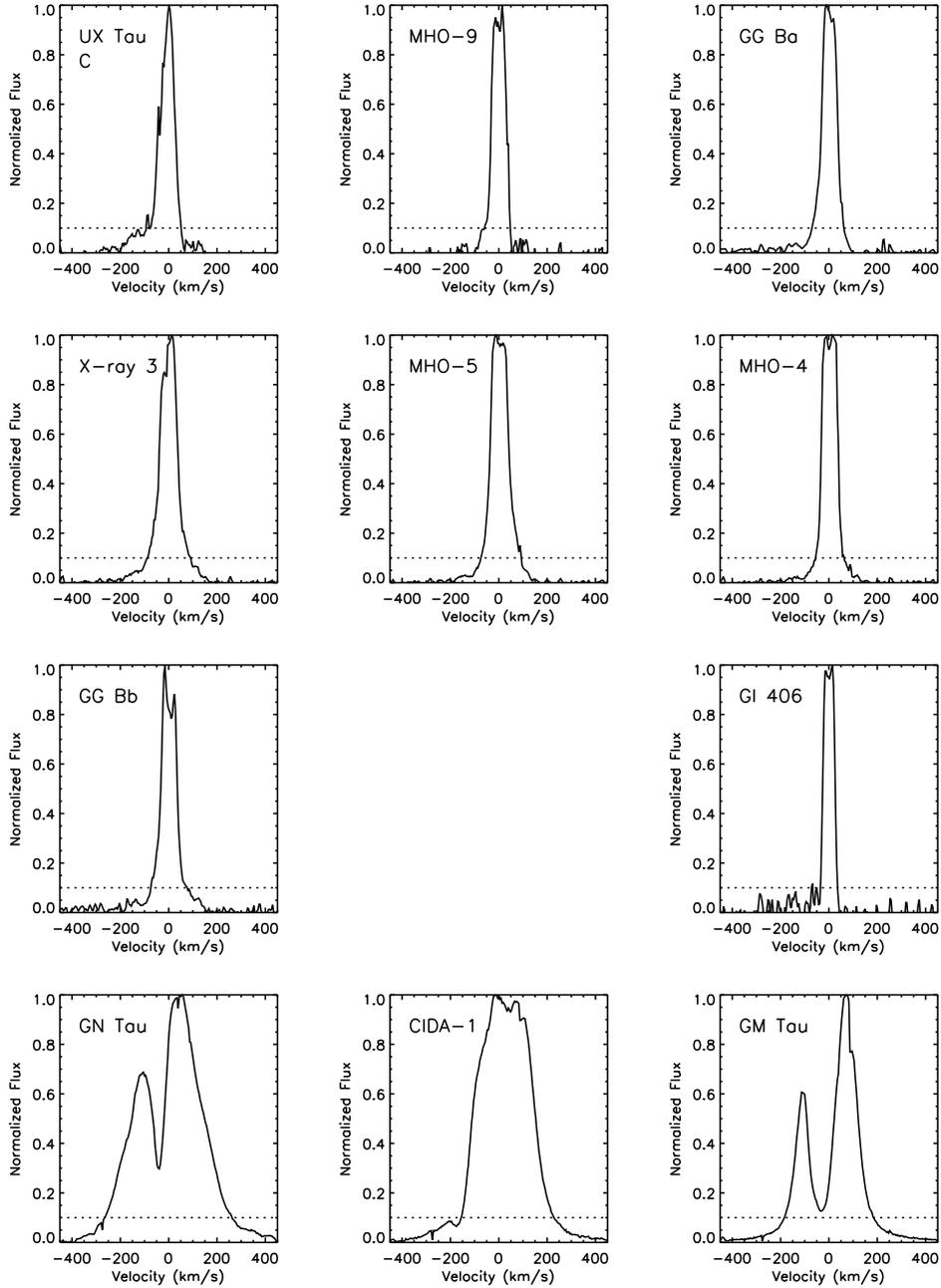}
\caption{H$\alpha$ velocity profiles for the LMTTs and the M6 dwarf Gl
406.  The profiles have been continuum subtracted and normalized to range
from 0.0 (continuum level) to 1.0 (emission peak level).  The profiles of
the LMTTs with no detected veiling (first 7 profiles) exhibit relatively
symmetric and narrow H$\alpha$ profiles, consistent with little or no
accretion.  In contrast, optically veiled LMTTs (last 3 profiles)
exhibit broad H$\alpha$ profiles, consistent with the predictions of the
magnetospheric accretion models.  The superimposed blue-shifted absorption
in these 3 profiles is indicative of a strong wind.
\label{fig_hatts}}
\end{figure}

\begin{figure}
\epsscale{0.80}
\plotone{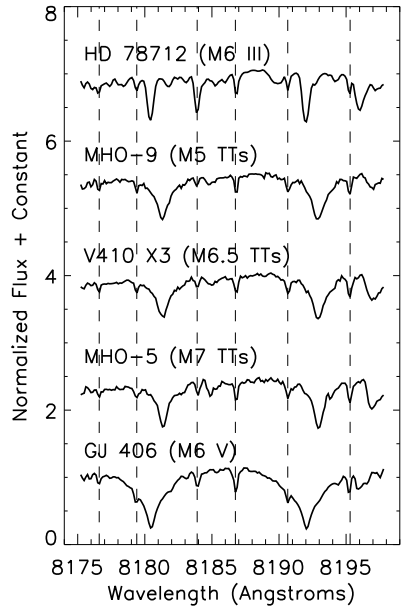}
\caption{
The pressure sensitive Na I absorption profiles for an M6 giant, 
an M6 dwarf, and 3 mid-M T Tauri stars.  The spectra are shown at their
observed velocities for easy identification of telluric features
(\textit{dashed lines}), which are not removed.  The Na I wing strengths of
the T Tauri spectra are intermediate between that of the narrow giant and
strongly broadened dwarf, and are consistent with an intermediate surface
gravity.
\label{fig_na}}
\end{figure} 

\begin{figure}
\epsscale{0.80}
\plotone{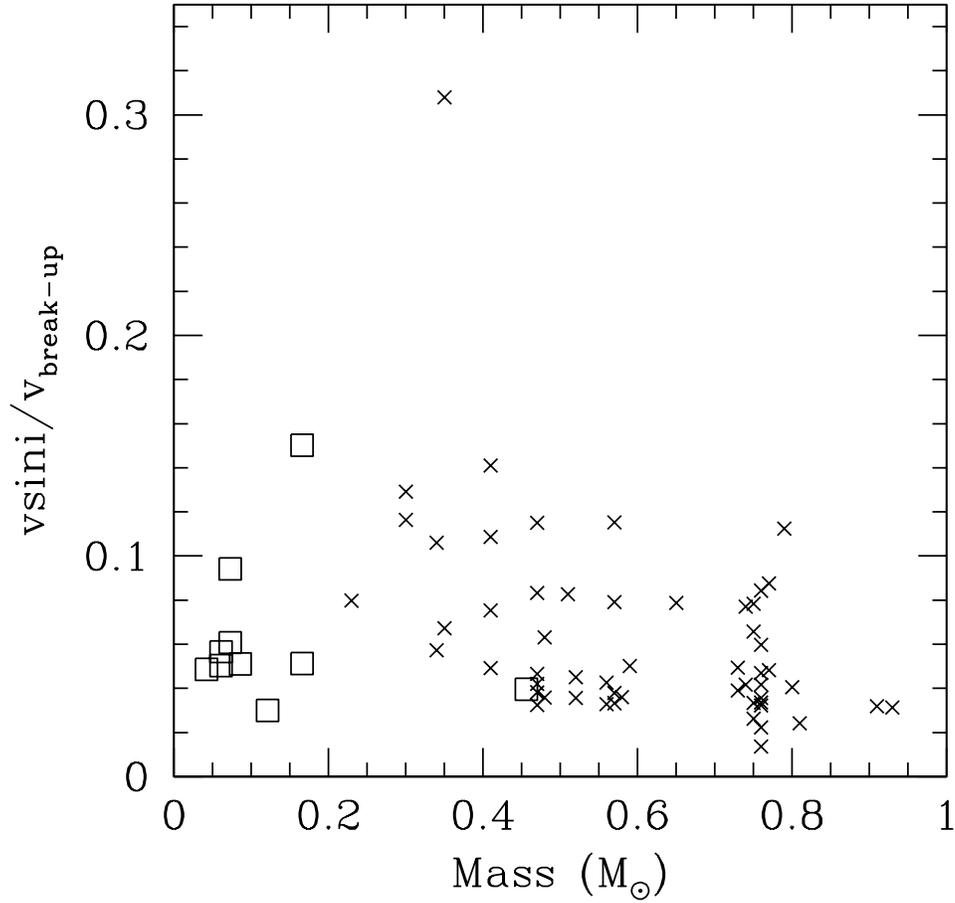}
\caption{Rotational velocities ($v$sin$i$) expressed as a fraction of the
break-up velocity plotted versus stellar mass.  The \textit{large squares}
represent new measurements presented here.  The \textit{small x's} are
rotational values for Taurus stars listed in \citet{cb00}.
\label{rot_mass}}
\end{figure} 

\begin{figure}
\epsscale{0.80}
\plotone{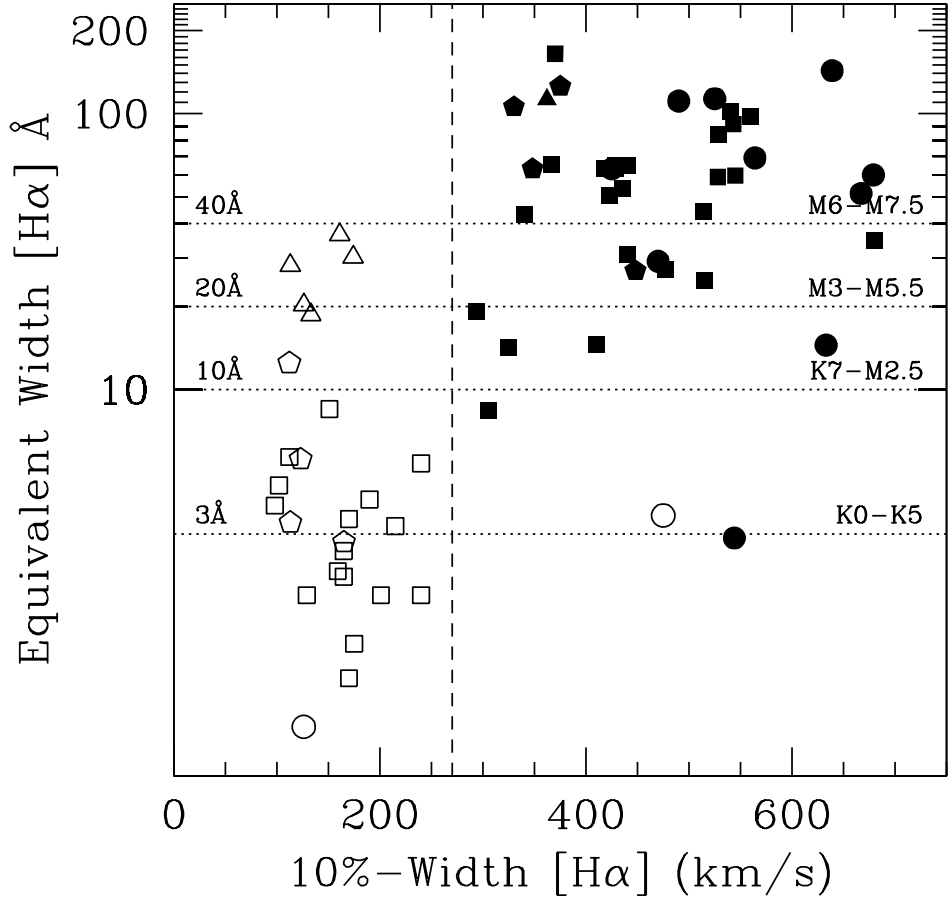}
\caption{Equivalent widths of H$\alpha$ versus 10\%-widths of H$\alpha$
for T Tauri stars.  The 10\%-widths of optically veiled T Tauri stars
(\textit{filled symbols}) are distinctively larger ($\gtrsim 270$)
than the 10\%-widths of non-optically veiled T Tauri stars
(\textit{unfilled symbols}), independent of spectral type.  Optically
veiled and non-optically veiled stars can be distinguished based on
EW[H$\alpha$], if the spectral type dependence is accounted for.  T Tauri
stars are usually veiled if the EW[H$\alpha$] $\ge$ 3 \AA\, for K0-K5 stars
(\textit{circles}), EW[H$\alpha$] $\ge$ 10 \AA\, for K7-M2.5 stars
(\textit{squares}), EW[H$\alpha$] $\ge$ 20 \AA\, for M3-M5.5 stars
(\textit{pentagons}), EW[H$\alpha$] $\ge$ 40 \AA\, for M6-M7.5 stars
(\textit{triangles}).  Measurements were extracted from data presented in
\citet{heg95}, \citet{ab00}, \citet{muzerolle00b}, and this work. 
\label{width}}
\end{figure}

\begin{figure}
\epsscale{0.80}
\plotone{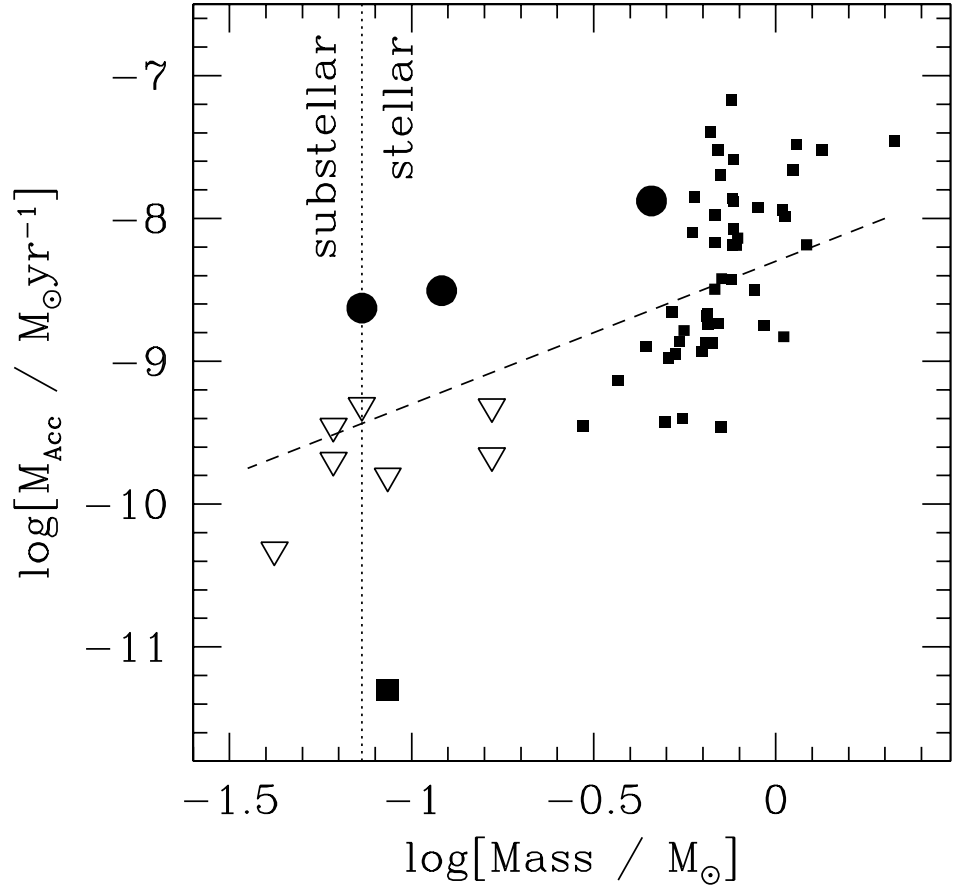}
\caption{Mass accretion rates versus stellar mass.  The large
\textit{circles} and \textit{triangles} are measurements presented here;
the \textit{triangles} are upper limits.  The \textit{large square} shows
the mass accretion rate derived for V410 Tau Anon 13 from its H$\alpha$
profile \citep{muzerolle00a}.  The \textit{small squares} are mass
accretion rates for classical T Tauri stars in Taurus from \citet{wg01}.
The mass accretion rate may be mass dependent.  The function form $\dot{M}
\propto M$ (\textit{dashed line}) is shown form comparison.
\label{mass_acc}}
\end{figure}

\newpage
\begin{deluxetable}{lcclllc}
\tabletypesize{\small}
\tablecaption{Spectroscopic Properties\label{tab_spec}}
\tablewidth{0pt}
\tablehead{ \colhead{}
& \colhead{EW[Li\,I]}
& \colhead{EW[H$\alpha$]}
& \colhead{}
& \colhead{Prev}
& \colhead{RV} 
& \colhead{$v$sin$i$} \\
\colhead{Star}
& \colhead{(\AA)}
& \colhead{(\AA)}
& \colhead{SpTp}
& \colhead{SpTp}
& \colhead{(km/s)}
& \colhead{(km/s)}
}
\startdata
GN Tau		&0.39	&-59.0	&M2.5	&C	&$16.7 \pm 1.3$	& $10.0 \pm 2.7$ \\
UX Tau C	&0.54	&-5.6	&M5	&M5:	&$14.7$		& $29 \pm 9$ \\
MHO-9		&0.54	&-3.3	&M5	&M4	&$16.2 \pm 0.8$	& $9.9 \pm 1.9$ \\
CIDA-1		&0.25	&-126	&M5.5	&C	&$16.2 \pm 0.7$	& $5.3 \pm 2.6$ \\
GG Tau Ba	&0.56	&-20.3	&M6	&M5	&$16.8 \pm 0.7$	& $8.1 \pm 0.9$ \\
GM Tau		&0.18	&-113	&M6.5	&C	&$15.3 \pm 1.8$	& $9.1 \pm 4.1$ \\
V410 X-ray 3	&0.46	&-30.2	&M6.5	&M6	&$17.9 \pm 1.2$	& $14.1 \pm 2.5$ \\
MHO-5		&0.53	&-36.4	&M7	&M6.5	&$20.8 \pm 0.7$ & $8.0 \pm 1.0$ \\
MHO-4		&0.45	&-28.2	&M7	&M6.5	&$19.1 \pm 0.4$	& $7.1 \pm 1.4$ \\
GG Tau Bb	&0.60	&-18.7	&M7.5	&M7	&$17.1 \pm 1.0$	& $6.6 \pm 2.0$ \\
\enddata
\tablecomments{Previous spectral types are from \citet{bm95},
\citet{briceno93, briceno98}, \citet{luhman98} and \citet{wgrs99}. }
\end{deluxetable}

\newpage
\begin{deluxetable}{llllll}
\tablecaption{Continuum Excesses \label{tab_excess}}
\small
\tablewidth{0pt}
\tablehead{
\colhead{Star}
& \colhead{r$_{6500}$}
& \colhead{r$_{7500}$}
& \colhead{r$_{8000}$}
& \colhead{r$_{8400}$}
& \colhead{template}
}
\startdata
GN Tau 		& $1.16 \pm 0.51$ & $0.58 \pm 0.50$ & 
	$0.76 \pm 0.55$ & $0.71 \pm 0.50$ & M2V \\
UX Tau C 	& $<0.11$ & $<0.27$ & $<0.26$ & $<0.14$ & $<$M4V+M6V$>$ \\
MHO-9	 	& $<0.05$ & $<0.42$ & $<0.11$ & $<0.13$ & $<$M4V+M6V$>$ \\
CIDA-1	 	& $1.67 \pm 0.31$ & $1.11 \pm 0.51$ & 
	$0.68 \pm 0.13$ & $0.69 \pm 0.28$	& $<$M4V+M6V$>$ \\
GG Tau Ba 	& $<0.06$ & $<0.09$ & $<0.15$ & $<0.19$ &  $<$M6V+M6III$>$ \\
GM Tau		& $2.29 \pm 0.43$ & $0.96 \pm 0.12$ &
	$1.10 \pm 0.32$ & $1.04 \pm 0.36$ & $<$M6.5V+M6.5III$>$ \\
V410 X-ray 3	&$<0.22$ & $<0.14$ & $<0.08$ & $<0.19$ & $<$M6V+M6III+M7V+M7III$>$ \\
MHO-5		& $<0.11$ & $<0.14$ & $<0.08$ & $<0.13$ & $<$M7V+M7III$>$ \\
MHO-4		& $<0.19$ & $<0.09$ & $<0.09$ & $<0.13$ & $<$M7V+M7III$>$ \\
GG Tau Bb	& $<0.11$ & $<0.08$ & $<0.08$ & $<0.15$ & $<$M7V+M7III+M8V+M8III$>$ \\
\enddata
\end{deluxetable}

\newpage
\begin{deluxetable}{llccccc}
\tablecaption{Stellar and Circumstellar Properties \label{tab_prop}}
\small
\tablewidth{0pt}
\tablehead{ \colhead{}
& \colhead{HIRES}
& \colhead{}
& \colhead{3 Myr} 
& \colhead{log($\dot{\textrm{M}}$)} \\

\colhead{Star}
& \colhead{SpTp}
& \colhead{T$_\textrm{eff}$}
& \colhead{Mass (M$_\odot$)}
& \colhead{(M$_\odot$/yr)}
}
\startdata
GN Tau	& M2.5	& 3488 & $0.456 \pm 0.088$ & $-7.9$ \\
UX Tau C& M5	& 3125 & $0.166 \pm 0.047$ & $<-9.3$ \\
MHO-9	& M5	& 3125 & $0.166 \pm 0.047$ & $<-9.7$ \\
CIDA-1	& M5.5	& 3058 & $0.121 \pm 0.040$ & $-8.5$ \\
GG Tau Ba& M6	& 2990 & $0.086 \pm 0.024$ & $<-9.8$ \\
GM Tau	& M6.5	& 2940 & $0.073 \pm 0.013$ & $-8.6$ \\
V410 X-ray 3&M6.5&2940 & $0.073 \pm 0.013$ & $<-9.3$ \\
MHO-5	& M7	& 2890 & $0.061 \pm 0.016$ & $<-9.7$ \\
MHO-4	& M7	& 2890 & $0.061 \pm 0.016$ & $<-9.5$ \\
GG Tau Bb& M7.5	& 2720 & $0.042 \pm 0.015$ & $<-10.3$ \\
\enddata
\end{deluxetable}

\newpage
\begin{deluxetable}{lccccc}
\tablecaption{Comparison Standards}
\small
\tablewidth{0pt}
\tablehead{ \colhead{}
& \colhead{}
& \colhead{}
& \colhead{$v$sin$i$}
& \colhead{RV}
& \colhead{} \\
\colhead{Name}
& \colhead{SpTp}
& \colhead{Ref}
& \colhead{(km/s)}
& \colhead{(km/s)}
& \colhead{Ref}
}
\startdata
\multicolumn{6}{c}{\underline{Dwarf Stars}} \\
Gl 15A		& M2	& 1 & $<2.9$ & $11.430 \pm 0.044$ & 2, 3 \\
Gl 447		& M4	& 2 & $<2.0$ & $-31.471 \pm 0.036$ & 2, 3 \\
Gl 876		& M4+	& 2 & $<2.0$ & $-1.800 \pm 0.200$ & 2, 3 \\
Gl 406		& M6	& 4 & $<2.9$ & $19.175 \pm 0.110$ & 2, 3 \\
VB 8		& M7	& 5 & $<6.8$ & $14.5 \pm 1.0$ & 6 \\
LHS 2243	& M8	& 5 & $<4.4$ & \nodata & 7 \\
\multicolumn{6}{c}{\underline{Giant Stars}} \\
SAO 63349	& M3	& 4 & $<11.6$	& $-43.5 \pm 1.5$	& 7 \\
HD 92620	& M3.5	& 1 & $<11.1$	& $13.2 \pm 1.4$	& 7 \\
HD 112264	& M5-	& 1 & $<10.3$ 	& $20.3 \pm 1.0$	& 7 \\
HD 94705	& M5.5	& 1 & $<10.2$ 	& $-10.5 \pm 0.9$	& 7 \\
HD 78712	& M6	& 8 & $<12.4$ 	& $11.9 \pm 0.9$	& 7 \\
HD 69243	& M7	& 8 & $<11.6$ 	& $31.8 \pm 1.1$	& 7 \\
HD 84748	& M8	& 8 & $<14.8$	& $3.6 \pm 2.5$		& 7 \\
\enddata
\tablecomments{The $v$sin$i$ measurements for the giant stars are
determined using dwarf templates and are therefore conservative upper
limits (see 3.1.2).}
\tablerefs{
(1) \citet{km89};
(2) \citet{delfosse98};
(3) Xavier Delfosse, priv. comm.;
(4) \citet{kirkpatrick91};
(5) \citet{kirkpatrick95};
(6) \citet{tr98};
(7) this study;
(8) \citet{hj82}.
}
\end{deluxetable}

\end{document}